\documentclass{jpsj3}
\usepackage{txfonts}
\usepackage{comment}
\usepackage{bm}
\usepackage{url}
\usepackage{color}
\usepackage{graphicx}

\title{
Structural Changes and Percolation Transition in Networks after Aging Processes
}

\author{Ryuho Sekikawa and Hiroshi Watanabe\thanks{hwatanabe@appi.keio.ac.jp}}
\inst{Department of Applied Physics and Physico-Informatics, Faculty of Science and Technology, Keio University, Yokohama 223-8522, Japan} 

\abst{
In social networking services, users constantly change, and the network structure changes simultaneously. As the network structure changes, so does the word-of-mouth within it. To study how information transfer on the network changes with the aging of the network, we investigated the relation of the structure and the percolation in the aged networks. We first prepared the Bianconi-Barab\'{a}si model as the initial network and observed the time evolution of its properties by repeatedly deleting and adding nodes. We introduced two tunable parameters, the deleting parameter $\alpha$ and the adding parameter $\beta$, and observed the aging behavior for the various parameter sets. We found that the network did not reach its steady state depending on the parameters. We also found that when the network is stable, there is a parameter region where the degree distribution changes from power-law to exponential decay. We performed the percolation analyses and found that the behavior of the percolation probability changes simultaneously with changes in the network's structure. This study is expected to help control network structure in steady-state aged networks.
}

\begin{document}
\maketitle

\section{Introduction}\label{sec:intro}

Network structures appear ubiquitously in the systems we see~\cite{Barabasi}. A network is defined by its components (nodes) and the interactions between them (edges). Networks in the real world are generally complex, but they are not random networks and have some universal properties. One of such properties is the small-worldness~\cite{Watts1998}. 
In a small-world network, arbitrary nodes are connected by relatively short path. It is known that networks in the real world are often also scale-free networks~\cite{BAmodel}. Scale-free networks have the property that a few nodes are connected to many other nodes. In contrast, many nodes have only a few edges, and the degree distribution asymptotically follows a power law. In real networks, such as the World Wide Web and metabolic networks, are known to be scale-free networks~\cite{Huberman1999, Faloutsos1999, Jeong_2000, Adamic2000}. Many methods have been studied to construct networks that can reproduce the properties of real networks. The Watts-Strogatz (WS) model is one of such algorithms to generate a network. Although the WS model successfully reproduces a small-world nature, the network constructed with this algorithm is not scale-free. The Barab\'asi Albert (BA) model based on preferential selection growth is widely known as an algorithm for generating scale-free networks~\cite{BAmodel}. In the BA model, the network grows from a small number of initial nodes, and at each time step, new nodes with $m$ edges are added to the network. These edges are connected to existing nodes with probability proportional to the degree of the existing nodes. This preferential growth is a key to generating a scale-free network. As the extension of the BA model, the Bianconi-Barab\'asi (BB) model was proposed~\cite{Bianconi2001}. In the BB model, each node is assigned a value called fitness. A new node is connected to an existing node with a probability proportional to product of degree and fitness. Therefore, nodes with a higher fitness are connected to more nodes with a higher probability. In the BA model, nodes added later cannot acquire more edges than those already in the network. However, in the BB model, nodes added later may also increase the possibility of gaining more edges than the nodes already existing in the network with the introduction of the fitness parameter.

These network generation models focus on how networks grow. On the other hand, as time passes after the network is constructed, nodes and edges may lose their activity. This effect is called aging~\cite{Amaral2000}. The impact of aging on the network has been actively investigated. Dorogovtsev and Mendes studied the evolution of networks with site aging, where the probability of connecting to older nodes decreases as $t^{-\alpha}$, where $t$ is the age of the node, i.e., the time since it was added to the network~\cite{Dorogovtsev2000}. They found that the shape of the degree distribution depends on the parameter $\alpha$. When $\alpha < 1$, the degree distribution follows the power-law, i.e., the network is scale-free, but the degree distribution becomes exponential for $\alpha > 1$.
Zhu \textit{et al.}~studied the effect of aging on network structures by introducing aging factors like exponential and power-law decay into evolving models~\cite{Zhu2003}. Aging changes network properties significantly, such as clustering, hierarchy, degree correlation, and average node distance, often leading to structural transformations. Moore \textit{et al.} obtained exact solutions of the structure of the network with addition and deletion of nodes~\cite{Moore}. The role of elementary processes such as the addition and deletion of nodes and edges on network evolution were summarized by Ghoshal \textit{et al}~\cite{Ghoshal2013}.

Although these studies have clarified the network evolutions with node/edge addition and deletion, it is not clear how information transfer on the network changes as the network structure changes. To investigate how the information transfer changes as the network evolves, we studied the relation between network structures and percolation transitions. To simulate aging process on a network, we introduced two tunable parameters for addition and deletion of nodes. We observed the degree distribution after the network reached its steady state. We also performed a percolation analysis to examine how information diffuses across the aged networks.

The rest of the paper is organized as follows. In section~\ref{sec:method}, we describe the method. The results are described in Sec.~\ref{sec:results}. Section~\ref{sec:summary} is devoted to the summary and discussion.

\section{Method}\label{sec:method}

\subsection{Construction and aging of networks}

\begin{figure}[htbp]
    \centering
    \includegraphics[width=10cm]{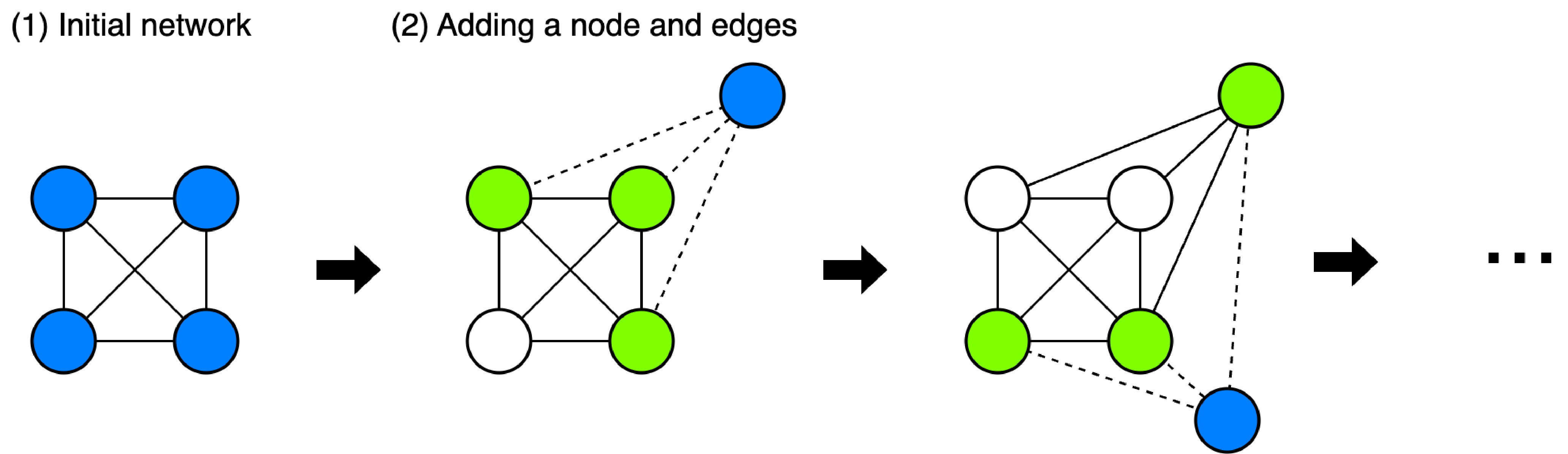}
    \caption{(Color online) Schematic illustration of the construction of the initial network. We adopt the BB model as the initial network. (1) We first prepare the complete graph with $m$ vertices. (2) Then, we add a node with $m$ vertices. The edge connecting the newly added node to the existing node is chosen in proportion to the product of the number of edges and the fitness of the existing node.}
    \label{fig:BB}
\end{figure}

We adopted the BB model instead of the BA model to account for the possibility that later users can acquire a large number of followers. We first prepared the initial network with BB model. The construction of the BB model is illustrated in Fig.~\ref{fig:BB}. First, we prepare the complete network of size $m$ in which the nodes are fully connected. The degree of node $i$ is denoted by $k_i$, which is the number of connected nodes to the node $i$. Each node is assigned a value called the fitness. The fitness of the node $i$ is denoted by $\eta_i$. 

We add a new node $j$ with $m$ edges and a fitness $\eta_j$ to the network. The probability $\Pi_i$ to connect an existing node $i$ and the new node $j$ is given by
\begin{equation}
    \Pi_i = \frac{\eta_i k_i}{\sum_l \eta_l k_l}. \label{eq:ba_model}
\end{equation}
As indicated in the above equation, the node with higher fitness has a higher chance of acquiring new edges. The adding step is repeated until the number of nodes reaches $N$. Finally, we obtain a scale-free network with $N$ nodes. 

\begin{figure}[htbp]
    \centering
    \includegraphics[width=10cm]{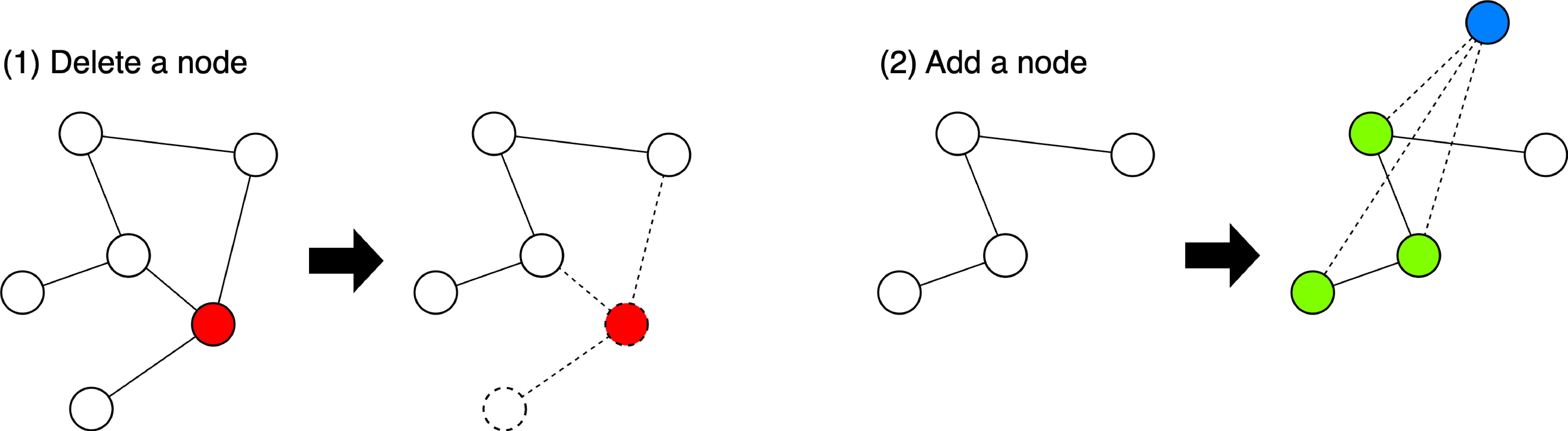}
    \caption{(Color online) Schematic illustration of the aging process after constructing the initial network. (1) Removing process. Delete a node $i$ with a probability proportional to $k_i^\alpha$ for its degree $k_i$. If this deletion results in the appearance of isolated nodes, those nodes are also deleted. (2) Adding process. Add nodes in a process similar to the BB model until the total number of nodes reaches $N$. Here, the probability of adding a node is modified by the parameter $\beta$. This process of deletion and addition is repeated until the network reaches a steady state.}
    \label{fig:aging}
\end{figure}

After the initial network is constructed, we repeatedly delete and add nodes to represent the aging process of the network (Fig.~\ref{fig:aging}). This study considers the deleting parameter $\alpha$ and adding parameter $\beta$. A node to be deleted is chosen in proportion to $k_i^\alpha$. When $\alpha$ is large, the node with more degree is more likely to be removed. If this removal results in isolated nodes with no edges, they are also deleted. After the node deletion, we add a new node $j$ with fitness $\eta_j$. The new node is connected to $m$ existing nodes. The probability of connecting a node $i$ and $j$ is given by
\begin{equation}
    \Pi_i = \frac{(\eta_{i} k_{i})^{\beta}}{\sum_{l} (\eta_{l} k_{l})^{\beta}},
\end{equation}
where $\beta$ is the adding parameter. This equation is reduced to Eq.~(\ref{eq:ba_model}) for $\beta = 1$. This adding process is repeated until the total number of nodes reaches $N$. Throughout this study, we assigned uniformly distributed random integers from $1$ to $100$. Here, we excluded the value of 0 from the fitness to avoid division by zero when the value of beta is negative.

We repeat the deleting and adding processes until the network reaches a steady state. Hereafter, a network that has reached a steady state will be referred to as an aged network. Throughout this study, we chose $m=4$ and $N=10000$. We independently generated 100 network samples for each parameter pair $(\alpha, \beta)$ and computed the statistical average of the observed quantities.

\subsection{Percolation analysis}

On social media, the diffusion of information among users, so-called word-of-mouth, is critical. While users with many followers are called influencers, and their statements greatly influence the networks, the influence of “ordinary influencers” who do not have a huge number of followers cannot be ignored~\cite{Bakshy2011}. How such word-of-mouth information diffuses on social media is expected to strongly depend on the distribution of the number of users' followers, i.e., the network's degree distribution. Therefore, we performed a percolation analysis to investigate how information diffuses on the aged network. The percolation analysis is a method used to study propagation and robustness within networks~\cite{Buldyrev_2010, LI20211}. In particular, complex networks such as scale-free networks are characterized by distinctive critical phases in bond percolation~\cite{1390282679444341248, PhysRevLett.103.168701}. 

On social media platforms, information is transmitted when one user posts information and another user forwards it to their followers in some way, for example, reposting on X (Twitter). In a network, nodes are users, edges connecting nodes are follower and followee relationships, and information is transmitted through edges so that the bond percolation can describe word-of-mouth in a network. Therefore, we considered the bond percolation as follows. 

After the aging process, we performed bond percolation analyses on the aged network. If the network was not simply connected, we extracted the largest connected component linked by edges and investigated bond percolation within that component. First, each edge in the network is defined as active with probability $p$ and inactive with probability $1-p$. Note that we interpret probability $p$ as the ease of conveying information, and therefore, unlike previous studies~\cite{Cohen2000}, $p$ is the activation probability rather than the delete probability. The activation of an edge corresponds to the user reposting some user's posts to the user's followers. Nodes connected by an activated edge are defined to the same cluster. We calculate the ratio of the largest cluster size to the system (the largest connected component), denoted by $P_\infty$ as a function of $p$. We averaged 100 independent samples at each $p$ to calculate $P_\infty(p)$. We then averaged $P_\infty(p)$ over 100 networks constructed independently with the same parameters.

\section{Results}\label{sec:results}

\subsection{Time evolutions of the aging processes}

In this study, we have prepared a scale-free network as the initial state, followed by repeated addition and deletion of nodes to represent the aging processes of the network. To confirm whether a network reaches a steady state, we observed the time evolutions of the average and variance of the degrees distribution. Suppose the $i$-th node has $k_i(t)$ vertices at step $t$. The degree average $\bar{k}$ and degrees variance $v(t)$ are defined by
$$
\begin{aligned}
    \bar{k} &\equiv \frac{1}{N} \sum_i k_i,\\
    v(t) &\equiv \frac{1}{N} \sum_i (k_i(t) - \bar{k})^2,
\end{aligned}
$$
where $N$ is the total number of nodes.

\begin{figure}[htbp]
    \centering
    \includegraphics[width=0.8\linewidth]{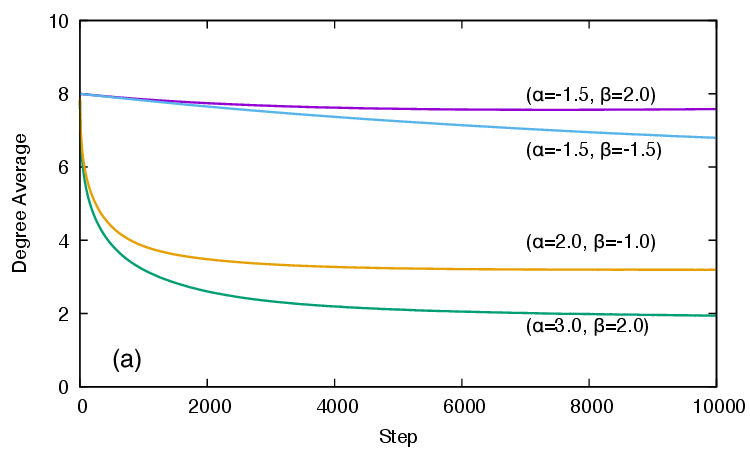}
    \includegraphics[width=0.8\linewidth]{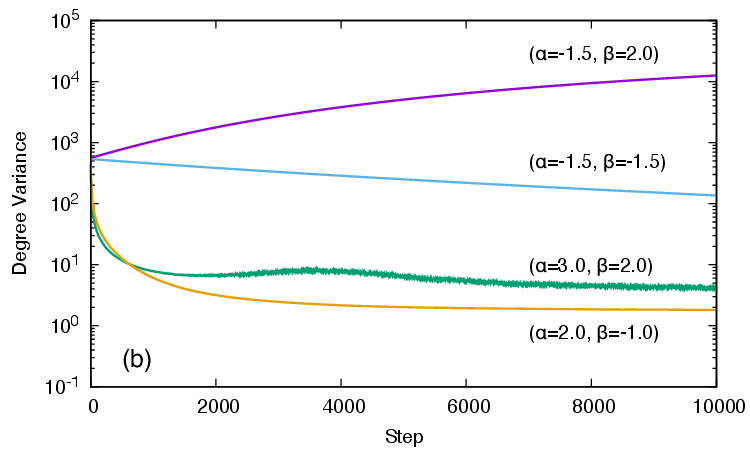}
    \caption{(Color online) Time evolutions of the aging processes. (a) Time evolutions of the average of degrees. (b) Time evolutions of the variance of degrees.}
    \label{fig:time_evolution}
\end{figure}

The time evolutions of the average of the degree distributions are shown in Fig.~\ref{fig:time_evolution}~(a). The initial values of the average $\bar{k}$ is $8$ which is the average of the degree distribution of the BB model. As the aging progressed, the averages appeared to converge to their respective parameter-dependent values. However, there were cases where the relaxation was extremely slow, for example at $(\alpha, \beta) = (-1.5, -1.5)$. We performed calculations with longer steps in this slow relaxation regime, but since the degree distribution and percolation behavior are almost unchanged. We, therefore, observed the degree distribution and percolation probability at the $10,000$ steps for comparison.

The time evolutions of the variance of the degree distributions are shown in Fig.~\ref{fig:time_evolution}~(b). The variance continued to increase with the aging process in some parameter region, such as at $(\alpha, \beta) = (-1.5, 2.0)$. We determined that the network did not reach steady state in these regions and did not include them in the phase diagrams shown in Fig.~\ref{fig:phasediagram}.

\subsection{Typical Configurations}

\begin{figure}[htbp]
    \centering
    \begin{picture}(14,185)
        \put(0,185){(a)}
    \end{picture}
    \includegraphics[width=0.45\linewidth]{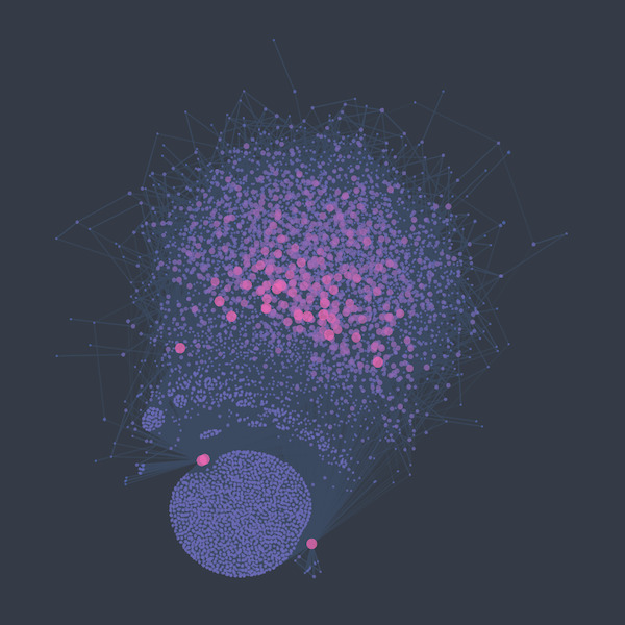}
    \begin{picture}(14,185)
        \put(0,185){(b)}
    \end{picture}
    \includegraphics[width=0.45\linewidth]{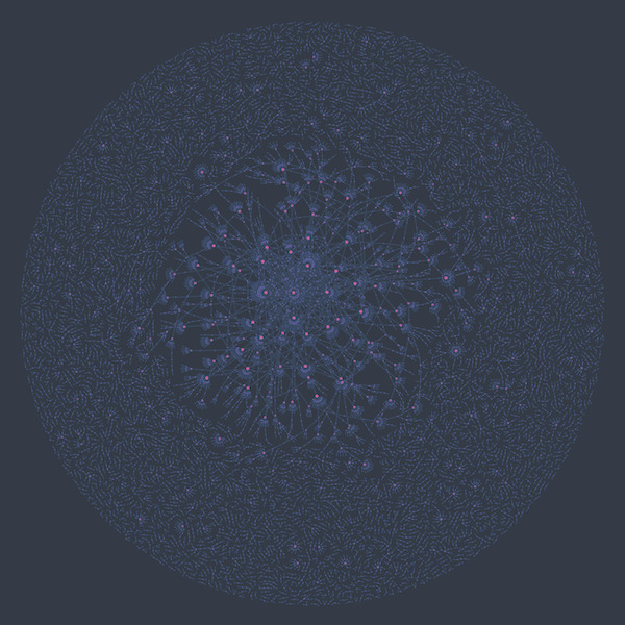}
    \begin{picture}(14,185)
        \put(0,185){(c)}
    \end{picture}
    \includegraphics[width=0.45\linewidth]{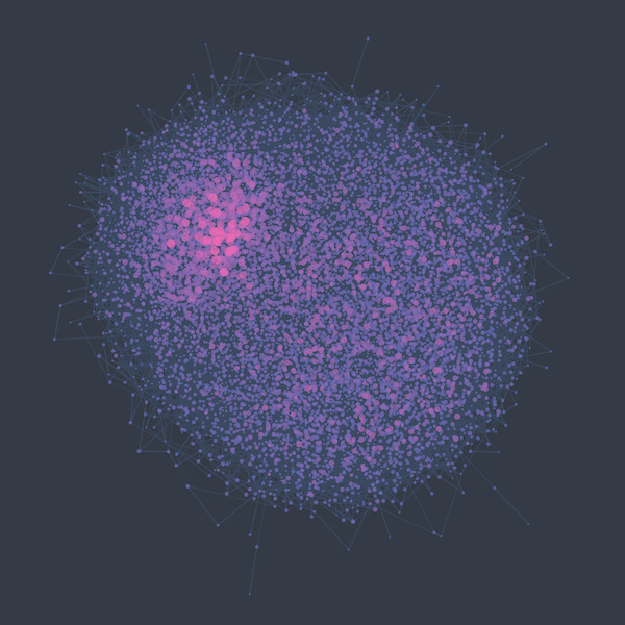}
    \begin{picture}(14,185)
        \put(0,185){(d)}
    \end{picture}
    \includegraphics[width=0.45\linewidth]{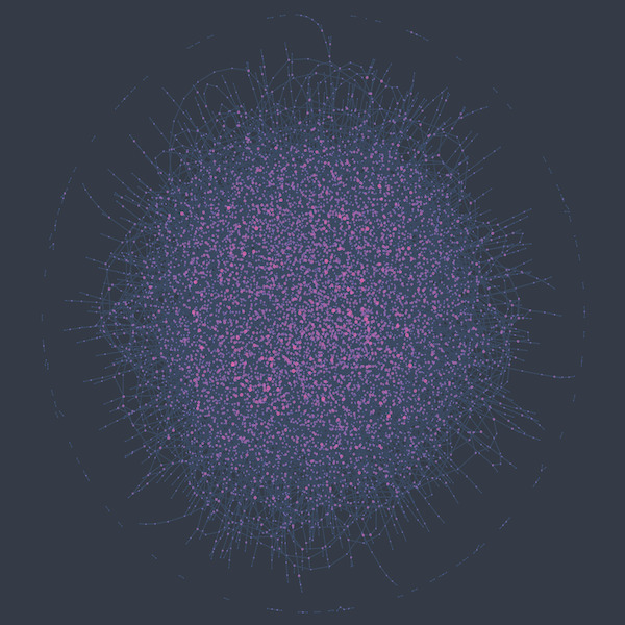}
    \caption{(Color online) Typical network structures generated by varying the parameters $\alpha$ and $\beta$. The values of $\alpha$ and $\beta$ for each panel are as follows: 
    (a) $(\alpha, \beta) = (-1.5, 2.0)$, 
    (b) $(\alpha, \beta) = (3.0, 2.5)$, 
    (c) $(\alpha, \beta) = (-1.5, -1.5)$, and 
    (d) $(\alpha, \beta) = (2.0, -1.0)$. 
}
    \label{fig:configurations}
\end{figure}

The typical configurations of the networks are shown in Fig~\ref{fig:configurations}. The networks are visualized by using Cosmograph\cite{Cosmograph}. As shown in Fig.~\ref{fig:configurations}~(a), in regions where both $\alpha$ and $\beta$ are small, most nodes belong to a single giant component. On the other hand, when $\alpha$ is increased, as shown in Fig.~\ref{fig:configurations}~(b), there is a giant component at the center of the network, and most of the nodes belong to it, but there are small connected components scattered in the surrounding area. When $\beta$ is increased, as depicted in Fig.~\ref{fig:configurations}~(c), although the giant component remains dominant, multiple connected components with non-negligible scales emerge. Furthermore, in regions where both $\alpha$ and $\beta$ are large, as shown in Fig.~\ref{fig:configurations}~(d), most nodes are dispersed into small connected components, indicating a structural transition. In some cases, the networks did not reach their steady states, which will be discussed later.

\subsection{Degree distribution}

\begin{figure}[htbp]
    \centering
    \includegraphics[width=0.8\linewidth]{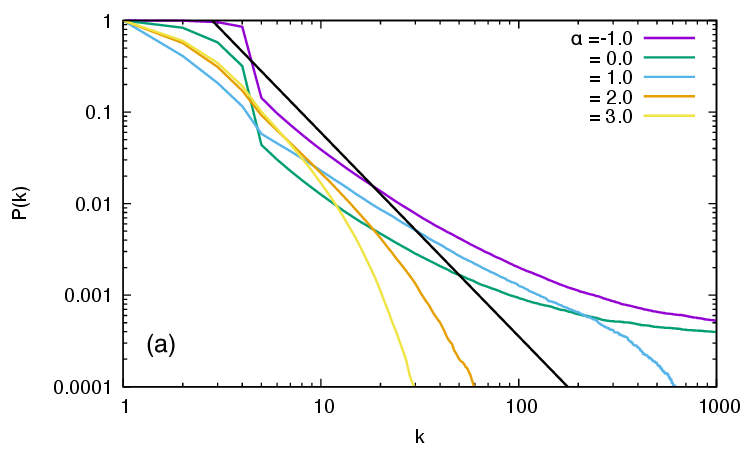}
    \includegraphics[width=0.8\linewidth]{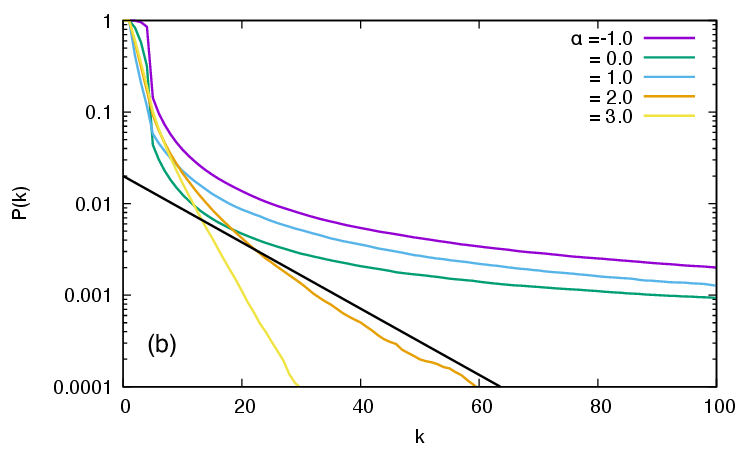}
    \caption{(Color online) The deletion parameter $\alpha$ dependence of the degree distribution. The value of the addition parameter is fixed to be $\beta = 2.0$. (a) The log-log plots. The solid line is the eye guide of the power-law decay. (b) The semi-log plots. The solid line is the eye guide of the exponential decay. As the value of $\alpha$ increases, the degree distribution changes from a power-law distribution to an exponential distribution.}
    \label{fig:alpha_dependence}
\end{figure}

\begin{figure}[htbp]
    \centering
    \includegraphics[width=0.8\linewidth]{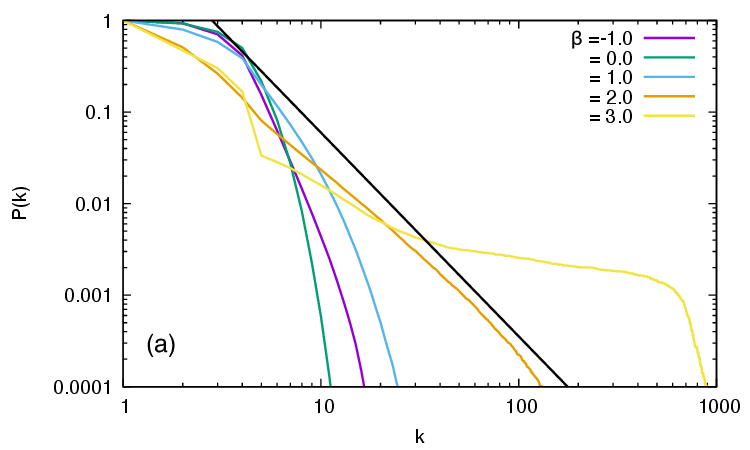}
    \includegraphics[width=0.8\linewidth]{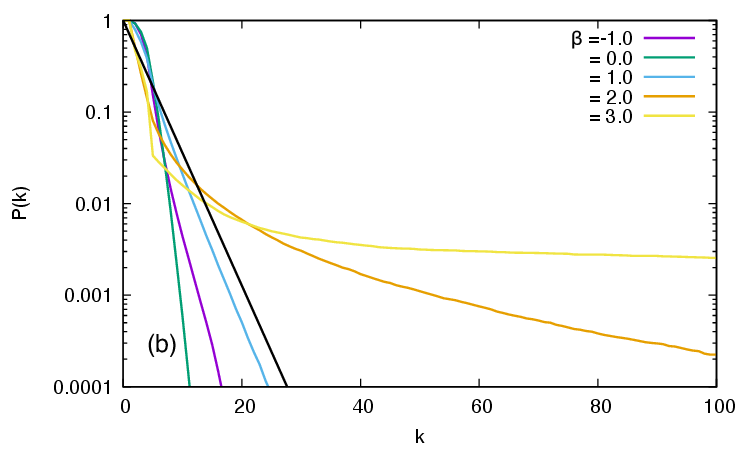}
    \caption{(Color online) The addition parameter $\beta$ dependence of the degree distribution. The value of the deletion parameter is fixed to be $\alpha = 1.5$. (a) The log-log plots. The solid line is the eye guide of the exponential decay. (b) The semi-log plots. The solid line is the eye guide of the power-law decay. As the value of $\beta$ increases, the degree distribution changes from a power-law distribution to an exponential distribution.}
    \label{fig:beta_dependence}
\end{figure}

We observed the complementary cumulative distribution function (CCDF) of the degree distribution $P(k)$, where $P(k)$ is the fraction of nodes with $k$ or fewer edges. The dependence of the degree distributions on the deleting parameter $\alpha$ for fixing the adding parameter $\beta$ to be $2.0$ are shown in Fig.~\ref{fig:alpha_dependence}. As the value of $\alpha$ increases, the degree distribution changes from power-law to exponential decay.

We also investigated the adding parameter $\beta$ dependence of the degree distribution by fixing the deleting parameter $\alpha$ to be $1.5$. The results are shown in Fig.~\ref{fig:beta_dependence}. While the degree distribution follows a power-law distribution for large values of the adding parameter, it exhibits exponential decay for small values. This behavior suggests that even when nodes are preferentially added to those that are not hubs, the scale-free nature of the network is disrupted.

To determine whether the degree distribution is power-law or exponential, we observes the following value,
$$
    \kappa = \frac{\left< k \right>}{\left<k^2\right>},
$$
where $\left< k \right>$ and $\left< k^2 \right>$ are the first and the second moment of the degree distribution. If the distribution is the exponential distribution with the exponent $\lambda$, then $\kappa = \lambda/2$. If the distribution follows the power-law with an exponent $\mu$, the value of $\kappa$ becomes zero when $\mu < 3$. Therefore, the value of $\kappa$ discriminates between exponential and power-law distributions. We adopt the threshold $\kappa = 0.1$, i.e., the network with $\kappa > 0.1$ was classified to be an exponential distribution.

\subsection{Percolation analysis}

\begin{figure}[htbp]
    \centering
    \includegraphics[width=0.8\linewidth]{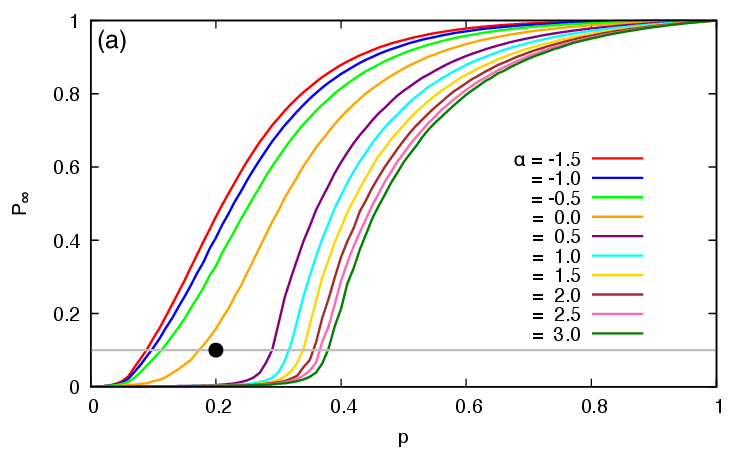}
    \includegraphics[width=0.8\linewidth]{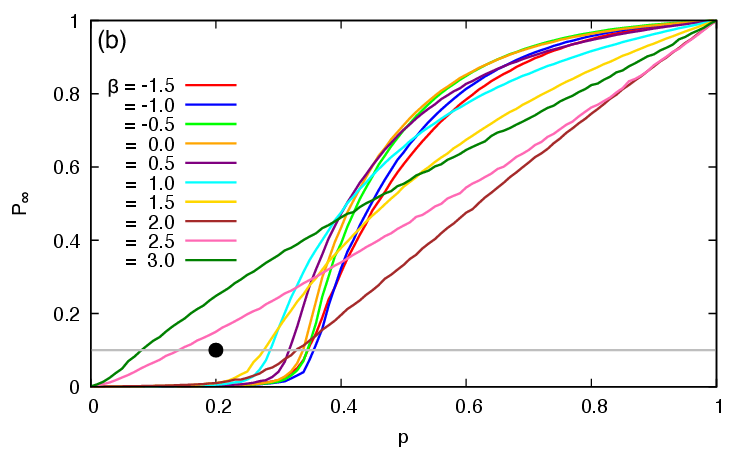}
    \caption{(Color online) Parameter dependence of percolation probability behavior. (a) $\alpha$-dependence of percolation probabilities with fixing $\beta = 0$. (b) $\beta$-dependence with fixing $\alpha = 2.0$. The position of the filled circle is $(p, P_\infty) = (0.2, 0.1)$ which was used as a reference point to determine whether the phase transition point is zero or finite.}
    \label{fig:percolation}
\end{figure}

We performed percolation analysis on the networks obtained after aging processes. If the network was not simply connected, we extracted the largest connected component linked by edges and investigated bond percolation within that component. The ratio of the volume of the largest cluster of the system (the largest connected component) to the active probability $p$ of the edge is shown in Fig.~\ref{fig:percolation}.

The percolation probabilities for different values of the deleting parameter $\alpha$, with the value of the adding parameter $\beta$ fixed to be $0$, are shown in Fig.~\ref{fig:percolation}~(a). Where the value of $\alpha$ changes from $0$ to $0.5$, the behavior of the percolation probability changes significantly. The deleting parameter dependence of the percolation probabilities are shown in Fig.~\ref{fig:percolation}~(b). The behavior of the percolation probability changes significantly when the value of $\beta$ changes from $2.0$ to $2.5$.

To capture these changes, we introduce thresholds $p_\mathrm{th}$ and $P_\mathrm{th}$ and define the percolation transition point to be finite if $P(p_\mathrm{th}) < P_\mathrm{th}$. Here, we chose $p_\mathrm{th} = 0.2$ and $P_\mathrm{th} = 0.1$ which is shown as the solid circle in Fig.~\ref{fig:percolation}.

\subsection{Phase Diagram}

\begin{figure}[htbp]
    \centering
    \includegraphics[width=0.8\linewidth]{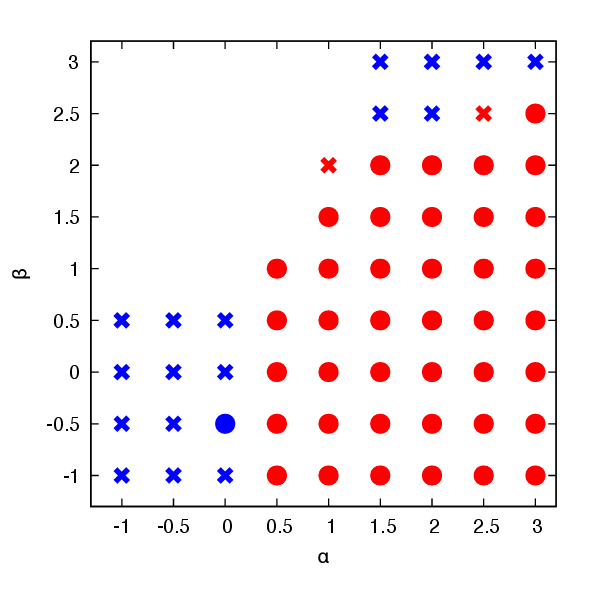}
    \caption{(Color online) Phase diagram of the network. Those with power-law distributions are marked with crosses, those with exponential distributions are marked with circles, those with finite percolation phase transitions are marked with red, and those with zero percolation phase transitions are marked with blue. The regions where the degree distribution is power-law and the phase transition point is zero almost overlap. Only networks that reached a steady state during the aging process were plotted.}
    \label{fig:phasediagram}
\end{figure}

We found that the degree distribution of the aged network changes from a power-law distribution to an exponential distribution depending on the parameters, and that the phase transition point of the percolation changes from zero to finite. We constructed a phase diagram from these results. The obtained phase diagram is shown in Fig.~\ref{fig:phasediagram}. This diagram classifies the characteristics of each network based on whether the degree distribution follows a power law and whether the percolation transition point is finite. The analysis reveals that the region where the degree distribution is exponential and the phase transition point of percolation is finite almost overlaps.

\section{Summary and Discussion}\label{sec:summary}

To investigate the network structure of a social media, we studied the aging processes of networks where nodes are continuously being deleted and added. We introduced two tunable parameters, $\alpha$ and $\beta$, to control deletions and additions of nodes. When the deleting parameter $\alpha$ is large, nodes with more edges are preferentially deleted. This represents a situation where users with many followers are more likely to leave the service. When the adding parameter $\beta$ is large, newly added nodes are more likely to be connected to nodes with more edges and fitness. This represents a situation in which new users to the service are more likely to follow the more popular users. We found that the scale-free nature of the network was destroyed in areas where $\alpha$ is large and $\beta$ is small. When one parameter was fixed, and the other parameter was varied, the degree distribution changed from a power-law distribution to an exponential distribution. This behavior is also seen in the aging process, where nodes lose activity over time~\cite{Dorogovtsev2000}. In our study, however, node activity does not change over time, and the structure of the network itself changes over time by repeating the aging procedure with fixed parameters. In this sense, we have investigated the aging process of the network itself, not the nodes or edges. 

On social media, the spread of word-of-mouth information is also essential. We investigated the percolation transition to consider the diffusion of the aged networks. The parameter regions where the degree distribution is exponential, i.e., where the network was not scale-free and where the percolation transition point is finite, were almost overlapped. The percolation on the network represents how some post goes viral. The fact that the percolation transition point is finite means that the post does not diffuse at all unless the impact of the post exceeds a certain threshold. As soon as the threshold is exceeded, it diffuses explosively. On the other hand, if the percolation transition point is zero, the diffusion of information will be more moderate and will spread according to the impact of the post. Our research suggests that the aging of the network can control word-of-mouth diffusion. While it is difficult to control leaving users, the properties of the network of joining users can be controlled to some extent. Possible applications include, for example, estimating the value of the deletion parameter by examining the nature of users who leave, determining the desired value of the addition parameter from the phase diagram, and deciding how to connect new users with existing users. One further issue is to examine the nature of social networks that have been in place long enough since launch to have reached a steady state and compare them to this study.

\begin{acknowledgment}
    The authors would like to thank N. Ito and Y. Murase for fruitful discussions. This research was supported by JSPS KAKENHI, Grant No. JP21K11923, and in part by RIKEN R-CCS HPC Computational Science Internship Program interns (2023). The computation was partly carried out using the facilities of the Supercomputer Center, Institute for Solid State Physics (ISSP), University of Tokyo.
\end{acknowledgment}

\appendix

\section{Effect of the fitness}

\begin{figure}[htbp]
    \centering
    \includegraphics[width=0.8\linewidth]{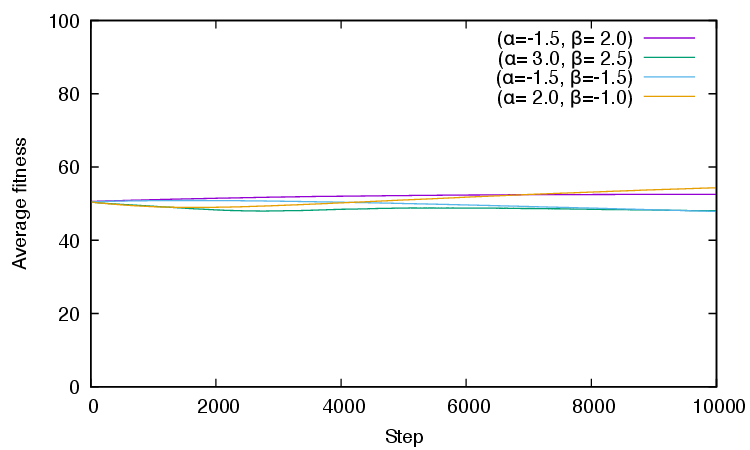}
    \caption{(Color online) Time evolutions of fitness.}
    \label{fig:fitness}
\end{figure}

In this study, we adopted the BB model to emulate the social networking services. The difference between the BA model and the BB model is the existence of the fitness. In order to investigate the effect of the fitness, we observed the time evolutions of the average fitness during aging processes. The time evolutions of the average fitness are shown in Fig.~\ref{fig:fitness}. As can be seen from the figure, the average degree of adaptation has changed little from the initial value of $50$.

To further investigate the effect of introducing fitness on the aging process, we performed a BA model-based aging simulation, i.e., when the fitness of the newly added node is fixed at $1$.

\begin{figure}[htbp]
    \centering
    \includegraphics[width=0.8\linewidth]{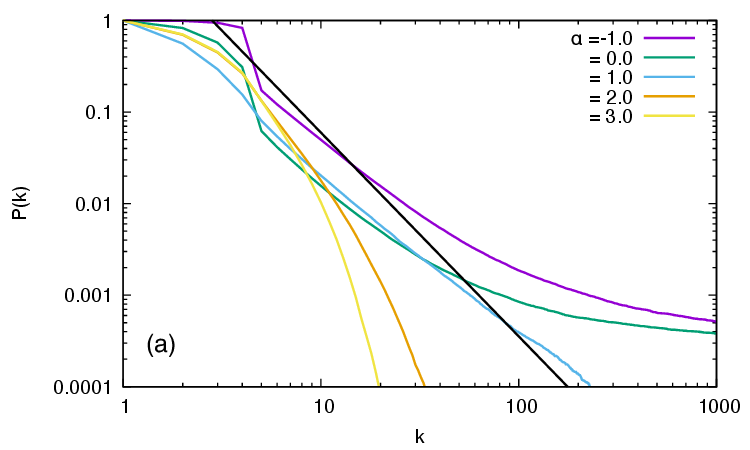}
    \includegraphics[width=0.8\linewidth]{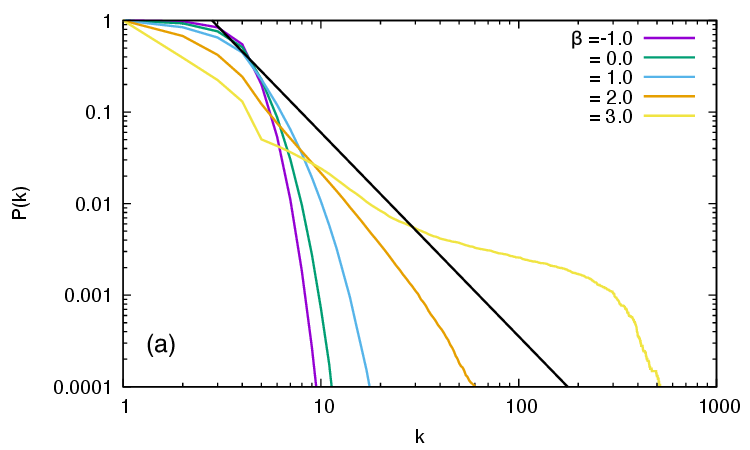}
    \caption{(Color online) Degree distributions of networks after the aging process with fixing fitness to be 1. The decimal logarithms are taken for both axes. (a) $\alpha$ dependence with $\beta = 2.0$. (b) $\beta$ dependence with $\alpha = 1.5$. The black solid line represents a power-law function as an eye-guide.}
    \label{fig:ba_model}
\end{figure}

The parameter dependencies of the degree distributions are shown in Fig.~\ref{fig:ba_model}. As the parameters changed, the degree distribution changed from a power-law distribution to an exponential decay, and its parameter domain was almost the same as in the BB model-based calculations. These results suggest that the influence of fitness was not significantly in the aging process.

\section{Finite-size effects on percolation}

\begin{figure}[htbp]
    \centering
    \includegraphics[width=0.49\linewidth]{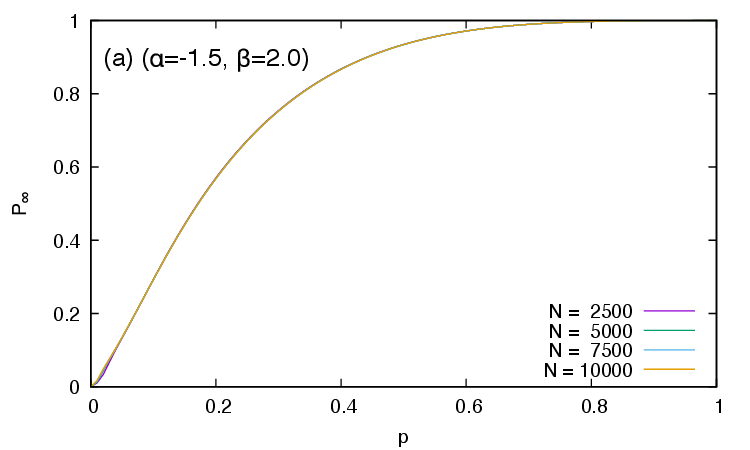}
    \includegraphics[width=0.49\linewidth]{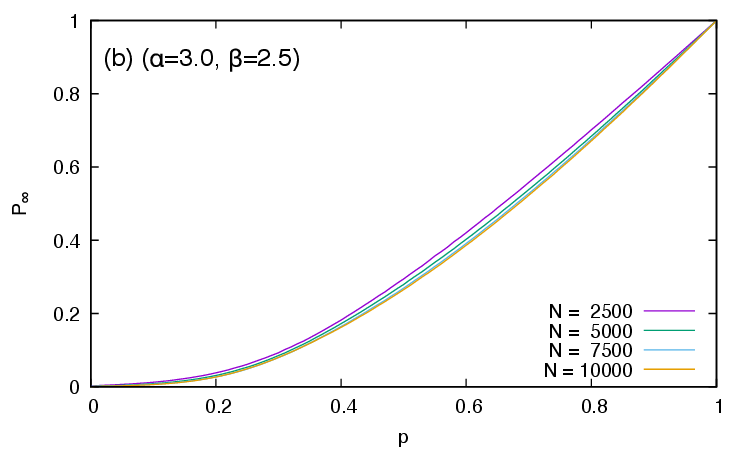}
    \includegraphics[width=0.49\linewidth]{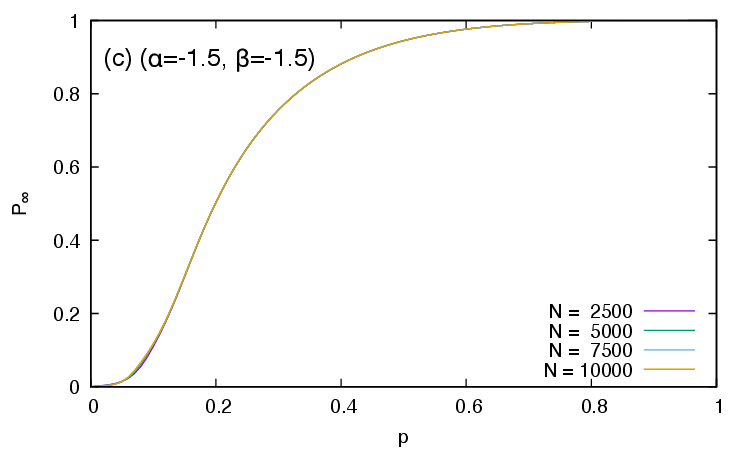}
    \includegraphics[width=0.49\linewidth]{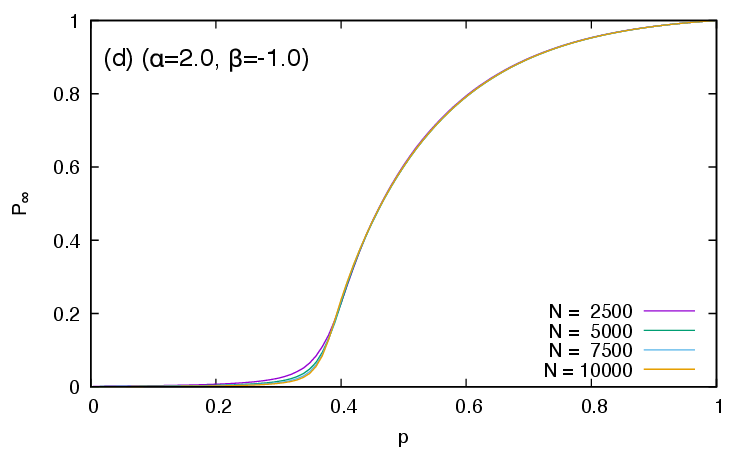}
    \caption{(Color online) Finite size effects on the percolation. (a) When $\alpha$ is small and $\beta$ is large. The variance increases monotonically in each step, and the system does not reach a steady state. (b) When $\alpha$ and $\beta$ are both large. (c) When $\alpha$ and $\beta$ are both small. (d) When $\alpha$ is large, and $\beta$ is small.}
    \label{fig:percolation_finite_size}
\end{figure}

We investigated the finite-size effects of percolation on the aged networks. The percolation probabilities of aged networks for the number of nodes $N=2500, 5000, 7500$, and $10000$ are shown in Fig~\ref{fig:percolation_finite_size}. The figures show no significant system size dependence was observed in most parameter regions. While the phase transition point was finite when the $N$ was small in regions where both the deleting and adding parameters are negatively large (Fig.~\ref{fig:percolation_finite_size}~(c)), the percolation probability quickly converged to some curve as the number of nodes increased and the phase transition point becomes zero. From these results, we conclude that finite size effects are sufficiently negligible at $N=10000$.

\bibliographystyle{jpsj}
\bibliography{reference}

\end{document}